\pdfoutput=1

\documentclass[%
twocolumn,
reprint,
superscriptaddress,
preprintnumbers,
nofootinbib,
showpacs,
aps,
prl
]{revtex4-2}

\makeatletter
\def\@hangfrom@section#1#2#3{\@hangfrom{#1#2}#3}
\def\@hangfroms@section#1#2{#1#2}
\makeatother

\usepackage[utf8]{inputenc}
\usepackage{graphicx, color}
\usepackage[svgnames,dvipsnames]{xcolor}
\usepackage{ragged2e}
\usepackage{dcolumn}
\usepackage{bm}
\usepackage{amsmath}
\usepackage{amsthm}
\usepackage{ascmac}
\usepackage{xparse} 
\usepackage{mathtools} 
\usepackage{blkarray} 
\usepackage{amscd,verbatim}
\usepackage{amssymb}
\usepackage[all]{xy}
\usepackage{tikz}
\usepackage{tikz-cd}
\usepackage{pgf,pgfplots}
\usepackage{enumitem}
\usepackage{slashed}
\usepackage{comment}
\usepackage{footmisc}
\usepackage{braket}
\usepackage{listofitems} 
\usetikzlibrary{arrows.meta} 
\usepackage[outline]{contour} 
\usepackage[colorlinks,citecolor=DarkGreen,linkcolor=FireBrick,linktoc=all]{hyperref}
\pgfplotsset{compat=newest}

\contourlength{1.4pt}

\usetikzlibrary{decorations.pathmorphing}

\tikzset{>=latex}

\renewcommand{\d}[0]{\mathrm{d}}

\allowdisplaybreaks

\begin{document}
\preprint{preprint}

\title{
Work distribution and fluctuation theorem in AdS/CFT
}

\author{Daichi Takeda}
\email{daichi.takeda@riken.jp}
\affiliation{iTHEMS, RIKEN, Wako, Saitama 351-0198, Japan}
\preprint{RIKEN-iTHEMS-Report-25}

\begin{abstract}
From the AdS/CFT dictionary, we derive a bulk dual of the work distribution defined by the two-point measurement on the boundary, yielding a bulk formulation of the Tasaki–Crooks fluctuation theorem.
We argue that this does not merely supply a holographic prescription for the work distribution; it encodes the mean energy change and fluctuations of bulk real-time dynamics associated with the two-point measurement.
\end{abstract}

\maketitle

\section{Introduction}
Thermodynamics provides a natural framework for studying the microscopic structure of gravity \cite{BardeenCarterHawking, HawkingHartle1st, Wald:1993nt, Bekenstein1}.
Macroscopic quantities such as the mass, angular momentum, and area of a black hole obey the first law, where the area plays the role of entropy \cite{HawkingHartle1st, Wald:1993nt, Bekenstein1}.
Accordingly, the second law corresponds to the area-increase theorem \cite{Hawking2nd,Bekenstein:1973ur, BekensteinGSL}, with many extensions and theoretical checks (see \cite{Wall:2009wm, Carlip:2014pma, Wall:2018ydq, Sarkar:2019xfd} for reviews).
This area-entropy relation was also corroborated by microstate counting in string theory \cite{Strominger:1996sh}.
Moreover, the Einstein equation can be interpreted as a local realization of the first law \cite{Jacobson:1995ab}.
Taken together, it is suggested that classical gravity should emerge as the coarse-grained behavior of some underlying microscopic structure.

To probe finer structure in thermodynamics, we must use nonequilibrium statistical mechanics, which accounts for fluctuations.
In mesoscopic systems, total entropy need not always increase; instead, the rarity of entropy-decreasing events is quantified by the (detailed) fluctuation theorem (FT).
The macroscopic second law then follows as a corollary of the FT.
FTs have been established for classical deterministic dynamics \cite{jarzynski2000hamiltonian}, stochastic Markovian processes \cite{CrooksFT}, and quantum systems \cite{Kurchan:2000rzb, Tasaki2000jarzynski} with subsequent generalizations.

FTs in gravitational settings have also been explored.
In \cite{Massar:1999wg}, the ratio of transition probabilities between two black hole states is characterized by their area difference.
Building on this, \cite{Iso:2010tz} derived a FT for the composite system of a black hole and matter in its exterior, and \cite{Iso:2011gb} verified its consistency using a scalar field on the Schwarzschild background.
FTs for matter and for local systems on curved spacetimes have been widely developed in \cite{Liu:2014hna, Basso:2023fua, Basso:2024yln, Cai:2024bgh, Costa:2025vqr}, and a FT in a semi-classical regime is also surveyed in terms of the gravitational algebra \cite{Cirafici:2024ccs, Cirafici:2024itu}.
Independently, \cite{Pourhassan:2021mhb, Pourhassan:2022sfk, Pourhassan:2022opb, Pourhassan:2022irk,Masood:2024yio} used Jarzynski equality for studying black hole thermodynamics.

Because quantum gravity remains unknown, derivations of FTs for dynamical gravity necessarily invoke approximations.
The formulation in \cite{Massar:1999wg}---on which \cite{Iso:2010tz} was based---employed a WKB approximation for the wavefunction together with a Born approximation for the detector–matter interaction.
Moreover, \cite{Iso:2010tz} analyzes a regime in which the total system is Markovian.
We should, however, note that those are not merely an idealization; \cite{Iso:2011gb} constructed an explicit model, derived the corresponding Langevin and Fokker--Planck equations, and thereby confirmed the FT of \cite{Iso:2010tz}.

Then, how can we derive FTs that hold in more general setups?
One natural route is the AdS/CFT (anti-de Sitter/conformal field theory) correspondence \cite{Maldacena:1997re, Gubser:1998bc, Witten:1998qj}.
For example, the origin of bulk entropy and the second law have been studied from the CFT standpoint.
In \cite{Engelhardt:2017aux, Engelhardt:2018kcs}, a coarse-grained entropy dual to the area of the bulk apparent horizon was proposed, providing a CFT interpretation of its area-increase law.
The CFT dual of the generalized entropy was derived in \cite{Chandrasekaran:2022eqq} using von Neumann algebra.
Conversely, \cite{Takeda:2024qbq, Shigemura:2024yeb} derived the second law by adopting a CFT entropy closer to equilibrium thermodynamics, and identified its bulk dual.

In this Letter, via the AdS/CFT dictionary, we derive a bulk expression of the Tasaki--Crooks quantum fluctuation theorem (TC) \cite{Tasaki2000jarzynski}.
The building blocks of TC are the free energy of the Gibbs ensemble and the work distribution $p(W)$ under the two-point measurement (TPM, reviewed later).
Since the bulk computation of the free energy is standard, our main target is a bulk representation of $p(W)$, which we will see realized as a gravitational Schwinger--Keldysh path integral \cite{Skenderis:2008dh, Skenderis:2008dg, Glorioso:2018mmw}.
At the end, we will evaluate $p(W)$ and the expectation value of the work in a concrete example, and verify TC in the bulk.

The main objective is to recast TC in the bulk.
On the other hand, we also present evidence that it should be viewed as a gravitational FT.
However, it remains for future work to identify the stochastic and/or quantum bulk dynamics dual to the boundary TPM.
Nevertheless, without prior knowledge of quantum gravity, holography allows us to compute $p(W)$, i.e., the statistics of the bulk energy change, which are constrained by TC.
In this sense, we expect this study to offer clues to quantum gravity.

\section{Tasaki--Crooks fluctuation theorem}
Here we introduce the TPM-based definition of the probability $p(W)$ \cite{Tasaki2000jarzynski} and its characteristic function \cite{Talkner2007fluctuation, Talkner2007tasaki}, and derive TC.
At the end, we state the setup in quantum field theories.

Let us first consider a quantum system, not a field theory.
The Hamiltonian is $H(t)$, and its $t$-dependence is given through protocol $\lambda$.
The eigenstate of $H(t)$ with eigenvalue $E_n^t$ is denoted by $\ket{E_n^t}$.
As the initial state, we prepare the Gibbs state $e^{-\beta H(0)}/Z(0)$ at inverse temperature $\beta$.
In the TPM, we first perform a projective measurement at $t=0$ onto $\ket{E_n^0}$, let the projected state evolve under $H(t)$, and at the final time $t=v$ perform a projective measurement onto $\ket{E_n^v}$.
The probability for this sequence of events is
\begin{align}\label{eq: F pmn}
  p_{m\to n} := |\braket{E_n^v|U|E_m^0}|^2\; \frac{e^{-\beta E_m^0}}{Z(0)}, 
\end{align}
where $U$ is defined as follows:
\begin{align}\label{eq: U}
    U := \mathrm{T}\exp\left[-i\int_0^v\d t\;H(t) \right]
    \quad
    (\mbox{$\mathrm{T}$: time-ordering}).
\end{align}

Next, following \cite{Andrieux2008quantum}, we introduce the time-reversal operator $\Theta$ and define the time-reversed evolution by
\begin{align}
	\tilde H(t) := \Theta H(v-t) \Theta^{\dagger},\qquad
	\tilde U := \Theta U^\dagger \Theta^{\dagger}.
\end{align}
Note that $\tilde H(t)$ shares the spectrum with $H(v-t)$, because the eigenvectors can be chosen as $\ket{\tilde E_n^t} = \Theta \ket{E_{n}^{v-t}}$.
Taking the initial state of the reverse process to be $e^{-\beta \tilde H(0)}/\tilde Z(0)$, the transition probability in the TPM is
\begin{align}
  \tilde p_{m\to n}& :=  |\braket{\tilde E_n^v|\tilde U|\tilde E_m^0}|^2\; \frac{e^{-\beta \tilde E_m^0}}{\tilde Z(0)}
  \nonumber\\
  &= |\braket{E_m^v|U|E_n^0}|^2\; \frac{e^{-\beta E_m^v}}{Z(v)}.
\end{align}

In the forward and reverse processes, we define the work distributions by
\begin{align}\label{eq: work dist}
  p(W) &:= \sum_{m,n}\delta (E_n^v - E_m^0 -W)p_{m\to n},\\
  \tilde p(W) &:=  \sum_{m,n}\delta (\tilde E_n^v - \tilde E_m^0 -W) \tilde p_{m\to n}.\label{eq: work dist tilde}
\end{align}
By using the Fourier representation of the delta function, we can write $p(W)$ as
\begin{align}
  p(W) &= \int \frac{\d u}{2\pi}\; e^{-i u W}G(u),\\
  G(u) &:= \mathrm{Tr}\left[U e^{-i(u - i\beta) H(0)}U^\dagger e^{i u H(v)} \right]/Z(0).
  \label{eq: G}
\end{align}
We call $G(u)$ the characteristic function.
By an analogous computation, we also find
\begin{align}
  \tilde p(W) &= \int \frac{\d u}{2\pi}\; e^{-i u W}\tilde G(u),\\
  \tilde G(u) &:= \mathrm{Tr}\left[U^\dagger e^{-i(u - i\beta) H(v)}U e^{i u H(0)} \right]/Z(v).\label{eq: G tilde}
\end{align}

Now, TC asserts the following:
\begin{align}\label{eq: TC}
  \tilde p(-W) = e^{-\beta (W-\Delta F)} p(W),
\end{align}
where we have defined the free energy difference $\Delta F$ as
\begin{align}
  e^{\beta \Delta F} := Z(0)/Z(v).
\end{align}
To prove it, we first show \cite{Talkner2007tasaki}
\begin{align}\label{eq: TC in G}
	\tilde G(-u + i\beta) = e^{-\beta \Delta F} G(u),
\end{align}
and then Fourier-transform this.

Taking the average of \eqref{eq: TC}, we obtain the Jarzynski equality \cite{Jarzynski1997nonequilibrium}:
\begin{align}
  \braket{e^{-\beta W}} = e^{-\beta \Delta F},
\end{align}
where $\braket{~}$ represents the average with $p(W)$.
From the Jensen inequality, we also have
\begin{align}\label{eq: Jarzynski}
  W_{\mathrm{diss}}: = \braket{W} - \Delta F \ge 0,
\end{align}
where $W_{\mathrm{diss}}$ is called dissipated work, and $\beta W_{\mathrm{diss}}$ is viewed as the entropy production.
If, after $t=v$, we couple the system to a heat bath at $\beta$ and let it thermalize, we have $\beta W_{\mathrm{diss}} = \Delta S + \beta Q$, where $\Delta S$ is the entropy change and $Q$ is the heat release to the bath after $t=v$.

In what follows, we set up the CFT framework in order to apply TC via AdS/CFT to gravity.
To make the discussion concrete, we consider the following action:
\begin{align}
  I_{\mathrm{tot}} = I_{\mathrm{CFT}} + \sum_I \int\d^d x\;\lambda^I(x) O_I(x).
\end{align}
Here $I_{\mathrm{CFT}}$ is the action of a holographic CFT, and each $O_I$ is a primary field.
The sum of the single-trace deformations must be a real scalar.
For $\Theta$, we can combine the time-reversal operator with parity and/or charge conjugation if necessary so that $I_{\mathrm{CFT}}$ is unchanged in the reverse process.

Strictly speaking, in quantum field theories, we have to circumvent the TPM to define $p(W)$, and a way is proposed in \cite{Ortega:2019etm} based on \cite{DornerQuantumWork, Mazzola:2014dvw}.
However, the formal expressions in this section remain unchanged and can be used as written.

\section{The bulk expression of TC}
\begin{figure}[t]
	\centering
	\includegraphics[width = 0.85 \columnwidth]{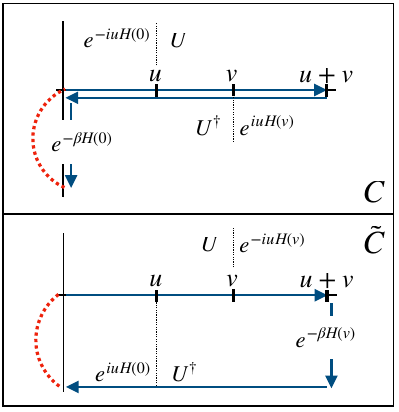}
	\caption{The closed time contours $C$ ($\tilde C$) for $G(u)$ ($\tilde G(u)$).
	The red dotted lines indicate periodic boundary conditions.
	}
	\label{fig: contours}
\end{figure}

Here we translate, via the AdS/CFT dictionary, \eqref{eq: TC in G} and \eqref{eq: TC} into the bulk.
Since \eqref{eq: TC in G} is built from $G(u)$, $\tilde G(u)$, and $\Delta F$, we rewrite these in bulk variables.

First, $\Delta F$ can be computed from the ratio of partition functions of the Gibbs ensembles, $Z(0)$ and $Z(v)$.
These partition functions admit bulk Euclidean path integral representations, approximated by Euclidean saddles in the large-$N$ limit.

\begin{figure}[t]
	\centering
	\includegraphics[width = \columnwidth]{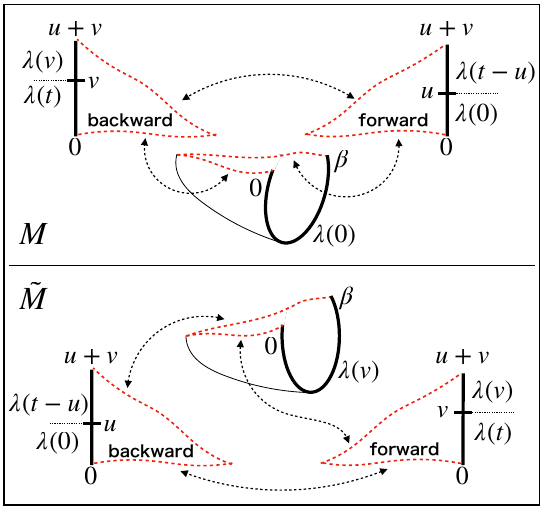}
	\caption{The bulk spacetime $M$ ($\tilde M$) to compute $G(u)$ ($\tilde G(u)$).
	Each consists of forward and backward Lorentzian segments and a Euclidean segment.
	While a black hole is drawn, a solution without horizon can also be a candidate saddle.}
	\label{fig: SvR}
\end{figure}

We next provide a bulk prescription for computing $G(u)$ and $\tilde G(u)$.
Notice that these quantities are written in the closed-time-path formalism (Fig.~\ref{fig: contours}).
This admits a Schwinger--Keldysh path integral representation.
Therefore, in the large-$N$ limit, we can use the prescription of \cite{Skenderis:2008dh, Skenderis:2008dg, vanRees:2009rw} to get:
\begin{align}\label{eq: bulk G}
  G(u) = e^{iS[\Phi; M]}/Z(0),\qquad
  \tilde G(u) = e^{iS[\Phi; \tilde M]}/Z(v).
\end{align}
Here $S$ is the bulk action, $\Phi$ collectively denotes the classical bulk fields, and $M$ ($\tilde M$) is a spacetime consisting of Lorentzian and Euclidean segments as in Fig.~\ref{fig: SvR}, whose boundary time runs along $C$ ($\tilde C$).
The backward segments enter in $S$ with a minus sign, and the Euclidean ones with a factor of $i$.
The field values on $\partial M$ are specified by $\lambda$ as usual; for example, a real scalar field asymptotes in Fefferman--Graham gauge to $z^{d-\Delta}\lambda$, where $\Delta$ is the scaling dimension of its dual CFT operator.
At each junction surface, an appropriate ``smoothness" boundary condition is imposed on the fields (see \cite{Skenderis:2008dh, Skenderis:2008dg} and End Matter).
From \eqref{eq: G} and \eqref{eq: G tilde}, $G(u)$ and $\tilde G(u)$ must be equal to $1$ when $\lambda(t)$ is constant, which has fixed the overall normalizations of \eqref{eq: bulk G}.
In End Matter, we will review in more detail with an example.

Now that \eqref{eq: bulk G} provides the bulk expression of \eqref{eq: G} and \eqref{eq: G tilde}, we can also compute \eqref{eq: work dist} and \eqref{eq: work dist tilde}.
We will later demonstrate \eqref{eq: TC} in a simple bulk model.

\section{What gravitational dynamics does $p(W)$ describe?}
Our main objective is to rewrite \eqref{eq: TC} in bulk variables, which has just been accomplished.
In doing so, we introduced a Schwinger--Keldysh spacetime $M$, but $p(W)$ must know a certain bulk dynamics ``dual to the TPM''.

To build intuition, it is helpful to first consider the meaning of $\braket{W}$.
On the CFT side, we find
\begin{align}
  \braket{W} = \mathrm{Tr}\left(H(v) U\rho_0 U^\dagger \right) - \mathrm{Tr}\left(H(0)\rho_0 \right),
\end{align}
with $\rho_0 = e^{-\beta H(0)}/Z(0)$.
This expression shows that, to obtain $\braket{W}$, one may forget TPM and instead consider the change in the energy expectation value under the unitary evolution $\rho_0\mapsto U \rho_0 U^\dagger$.

The bulk dual of this process is standard.
In the bulk, we prepare a stationary black hole with source $\lambda|_{t=0}$, corresponding to $\rho_0$.
Since $\lambda$ is time-dependent, the bulk fields are excited after $t=0$ through the boundary conditions.
We thus solve this initial-value problem (IVP).
The bulk dual of $\braket{W}$ is the change of the Brown--York (BY) energy \cite{Brown:1992br}, a definition of the bulk energy \cite{Balasubramanian:1999re}.
We will explicitly confirm this later in an example.

However, the ``dual of the TPM" itself would not be well described by a classical gravity.
If it could, a deterministic process would fix the BY energy, where a notion of work \textit{distribution} does not make sense.
In quantum gravity, a gravitational Euclidean path integral \cite{Hartle:1976tp} would be dual to $\rho_0$, and be analytically continued to Lorentzian.
This state then undergoes nontrivial quantum dynamics, in which $p(W)$ will be the distribution of some off-shell notion of BY energy.

We do not pursue further exploration of this bulk quantum process in this Letter.
It must be, however, emphasized that $p(W)$ and thus $\braket{W^n}$ are nevertheless computable in the bulk, as was demonstrated in the previous section.
In short, the large-$N$ analysis on the auxiliary manifold $M$ yields $p(W)$, which satisfies the fluctuation theorem, and indicates that the real-time bulk dynamics dual to the TPM fluctuates around the classical IVP trajectory.
Below, we will examine this further by using an example.

\section{An example: scalar field}
The above construction is non-perturbative in $\lambda$.
In what follows, we compute $p(W)$ to $\mathcal{O}(\lambda^2)$ and demonstrate \eqref{eq: TC} using a simple model in AdS$_3$/CFT$_2$.
For concreteness, we consider only a real scalar primary $O$ ($1<\Delta<2$) coupled to a source $\lambda$ with $\lambda(0)=\lambda(v)=0$ (i.e., $Z(0)=Z(v)$).
On the bulk side, we suppose the BTZ black hole to realize the dominant saddle of $Z(0)$.
The bulk dual of $O$ is a real scalar field $\Phi$, and a free theory suffices to $\mathcal{O}(\lambda^2)$:
\begin{align}\label{eq: bulk action}
    &S = -\frac{1}{2}\int \d^3 x \sqrt{-g} \left[g^{\mu\nu}\partial_\mu \Phi \partial_\nu \Phi +m^2\Phi^2\right] + S_\mathrm{ct},\\
    &g_{\mu\nu}\d x^\mu \d x^\nu = -(r^2 - r_+^2)\d t^2 +\frac{\d r^2}{r^2-r_+^2} +r^2 \d\theta^2.\label{eq: BTZ}
\end{align}
Here, $S_\mathrm{ct}$ is the counterterm proposed in \cite{deHaro:2000vlm}.
The horizon radius $r_+$ is related to $\beta$ via $r_+ \beta = 2\pi$.

At this order, no gravitational perturbation is required in $M$ and $\tilde M$, so one might worry whether we are really seeing a ``gravitational FT".
However, we will confirm below that $p(W)$ contains information about the metric solving the IVP, as well as fluctuations around it.

The above model predicts $p(W)$ to $\mathcal{O}(\lambda^2)$ as
\begin{align}\label{eq: ex p}
  p(W) &= (1-q)\delta(W) +  e^{\beta W/2}\; \frac{\pi(\Delta -1)r_+^{2\Delta - 2}}{\Gamma(\Delta)\Gamma(\Delta + 1)}\times\nonumber\\
  &\qquad \int\! \frac{\d k}{(2\pi)^2}\left|\lambda_{W,k} \Gamma(\gamma_{W,k}) \Gamma(\gamma_{W,-k})\right|^2,\\
  q &:= \int_{W\neq 0}\d W\; p(W),\\
\gamma_{\omega ,k} &:= \frac{\Delta}{2} + \frac{i}{2r_+}(\omega + k),
\end{align}
where $\lambda_{\omega,k}$ is the Fourier transform of $\lambda(t)$, and $\Gamma$ denotes the gamma function.
If $\theta$ is periodic, $k$ is discretized.
A similar computation reveals $\tilde p(W)=p(W)$.
Therefore, we can now easily confirm \eqref{eq: TC}.
End Matter shows the derivation of $p(W)$.

We can compute $\braket{W}$ from \eqref{eq: ex p} as
\begin{align}\label{eq: scalar average work}
  \braket{W} &\propto \beta^{2-2\Delta}\!\! \int\! \frac{\d \omega \d k}{(2\pi)^2}\left|\lambda_{\omega,k} \Gamma(\gamma_{\omega,k})\Gamma(\gamma_{\omega,-k})\right|^2\omega \sinh \frac{\beta \omega}{2},
\end{align}
where we have omitted an inessential positive coefficient.
The positivity of \eqref{eq: scalar average work} is a reflection of \eqref{eq: Jarzynski}.

As we have mentioned, $\braket{W}$ should also be calculated by directly solving the IVP from the BTZ background, with the metric backreaction included.
This problem has already been solved for a $\theta$-independent source $\lambda (t) = w(t)$ as Eq.~(4.15) of \cite{Shigemura:2024yeb}.
We find that the values agree; because a spatially homogeneous source entails a trivial volume divergence, we compare energy densities.
Therefore, this model confirms that $p(W)$ encodes the BY energy of the IVP trajectory.

The information provided by $p(W)$ is not exhausted by the \textit{mean} BY energy---higher moments are calculable.
In particular, one finds that the variance of $W$ is strictly positive already at $\mathcal{O}(\lambda^2)$.
Therefore, the bulk dynamics dual to the TPM admits fluctuations around the deterministic evolution of the IVP.

\section{Summary and discussions}
By applying the AdS/CFT dictionary, we expressed the work distribution $p(W)$ defined on the boundary in terms of bulk quantities.
For a cyclic protocol with a weak source, we explicitly computed $p(W)$ in the bulk and confirmed the theorem.
Moreover, we found that $\braket{W}$ agrees with the BY energy obtained from the metric perturbation in the IVP, implying that $p(W)$ actually encodes gravitational dynamics.
Not only that, the non-zero variance indicates that the bulk dynamics dual to the TPM admits fluctuations around the IVP trajectory.
In other words, a controlled classical calculation of $p(W)$ on the auxiliary manifold $M$ necessarily accesses information beyond the deterministic evolution of the IVP, and the corresponding dynamics is constrained by the fluctuation theorem established holographically.

While we have treated a single-trace deformation, the generalization to multi-trace deformations can be implemented as usual \cite{Aharony:2001pa, Witten:2001ua, Berkooz:2002ug, Sever:2002fk, Aharony:2005sh, Aharony:2006hz}.
Although we chose a scalar primary operator in the example, the theorem should likewise be explicitly verified for operators with charge or spin.
Also, the FT must hold even in the semiclassical regime and for higher orders in $\lambda$.

FTs come in various forms (see, e.g., \cite{Shiraishi2023introduction}).
From which FT can the second law mentioned in the introduction \cite{Engelhardt:2017aux, Engelhardt:2018kcs, Chandrasekaran:2022eqq, Takeda:2024qbq, Shigemura:2024yeb} be derived, or how does the present FT relate to those results?
Also, the FT of \cite{Iso:2010tz} may be reexamined from the viewpoint of holography.
Such a perspective will lead us to the extension of the generalized entropy and to an understanding the origin of its increase even beyond the semiclassical regime.

In particular, FT has also been studied for composite systems (see e.g., \cite{Esposito2009nonequilibrium}), so one possible direction is to apply it to composite holographic CFTs.
Such setups have been analyzed in \cite{Aharony:2006hz, Karch:2023wui, Geng:2023ynk}, and \cite{Shigemura:2024yeb} studied their thermodynamic aspects.
More recently, \cite{Karch:2025hof} investigated energy dissipation through the boundary.
As another direction, the Lindblad equation has also been implemented in holography \cite{Ishii:2025qpy}.
It may therefore be possible to survey holographic predictions of gravitational fluctuation theorems for open or composite systems with gravity.

Finally, we should aim to formulate the bulk dual of the TPM more precisely and to relate $p(W)$ to the bulk quantum process, which would be a kind of ``unraveling" of the classical trajectory of the IVP.
This will advance our understanding of quantum gravity and at the same time, a formulation beyond AdS may come into view.

\subsection{Acknowledgements}
I thank Takanori Ishii, Satoshi Iso, and Toshifumi Noumi for discussions.
My work is supported by RIKEN Special Postdoctoral Researchers Program.

\section{End Matter}
We first review a more concrete prescription for constructing $M$ in Fig.~\ref{fig: SvR} and the solution $\Phi$ on it (the reverse process is analogous).
We consider solving perturbatively in $\lambda(t)-\lambda(0)$.
In such a situation, we first fix the background field configuration from the stationary Euclidean saddle of $Z(0)$.
Its Killing vector selects bulk time slices, extended up to the bifurcation surface of the Killing horizon if it exists.
Analytically continuing this time coordinate, one can construct $M$ with the junction conditions automatically satisfied at this zeroth order.

Once the time coordinate and the gluing surfaces are fixed, we solve for the field perturbations subject to the boundary conditions on $\partial M$ determined by $\lambda(t)-\lambda(0)$, and also to the junction conditions on the gluing surfaces that we have fixed.
In principle, this procedure can be iterated to arbitrary order.
Although we have written \eqref{eq: bulk G} entirely classically, a semi-classical computation including quantum corrections of the perturbative fields should likewise be implementable in this way.

Based on the above strategy, let us carry out the detailed computation for the free scalar field in the main text.

When the source is switched off and the BTZ spacetime is the dominant saddle, the spacetime $M$ in Fig.~\ref{fig: SvR} is constructed from Euclidean and Lorentzian BTZ.
We take the gluing surfaces to lie at constant values of the Killing time $t$ of \eqref{eq: BTZ}.
This is the background on which we compute the perturbative solution of the scalar field.

Solving the Klein-Gordon equation on the metric \eqref{eq: BTZ}, we find a solution $f_{\omega,k}$ in Fourier space as
\begin{widetext}
\begin{align}
  f_{\omega, k}(r) &:= r^{\Delta - 2}\left(1 - \frac{r_+^2}{r^2} \right)^{i \omega/(2r_+)} \frac{\Gamma(\gamma_{\omega,k})\Gamma(\gamma_{\omega,-k})}{\Gamma(\Delta-1)\Gamma(\delta_{\omega,k})}\; {}_2F_1\left(\delta_{\omega,k}-\gamma_{\omega,k},\delta_{\omega,k}-\gamma_{\omega,-k},\delta_{\omega,k};1-\frac{r_+^2}{r^2} \right)\\
  \Bigl(\delta_{\omega,k} &:=\gamma_{\omega,k}+\gamma_{\omega,-k}+1-\Delta \Big).\nonumber
\end{align}
\end{widetext}
Here ${}_2F_1$ denotes the Gauss hypergeometric function.
The solution is normalized as $f_{\omega,k} \sim r^{\Delta-2}$ as $r\to \infty$, and it behaves as the outgoing mode near the horizon.
Another independent solution, which is normalizable, is then given by $N_{\omega,k} := f_{\omega,k}-f_{-\omega,k}$.

According to \cite{Skenderis:2008dh, Skenderis:2008dg}, the fields are glued across the forward, backward, and Euclidean segments subject to certain “smoothness” conditions.
First, the field values are continuous across each junction surface.
Second, the time derivatives satisfy the following conditions.
Let the scalar field on the respective segments be $\Phi_f$, $\Phi_b$, and $\Phi_e$, and $\tau \in [0,\beta]$ be the Euclidean-segment time.
We impose $\partial_t \Phi_f = \partial_t \Phi_b$ at $t_f = t_b = u + v$, $\partial_t \Phi_b = i \partial_\tau \Phi_e$ at $t_b = 0$ and $\tau=0$, and $\partial_t \Phi_f = i \partial_\tau \Phi_e$ at $t_f=0$ and $\tau = \beta$.
These conditions are required from the stationarity of the total action $iS_f - iS_b - S_e$.

From these and the asymptotic boundary conditions, the solution on $M$ is found to be
\begin{widetext}
\begin{align}
  \Phi_f &= \int\frac{\d\omega \d k}{(2\pi)^2}e^{-i\omega t + i k \theta}\lambda_{\omega,k}\left[e^{iu\omega}f_{-\omega,k}+(1-e^{ iu\omega})e^{-\beta\omega}n_{\omega}N_{\omega,k} \right],\\
  \Phi_b &= \int\frac{\d\omega \d k}{(2\pi)^2}e^{-i\omega t + i k \theta}\lambda_{\omega,k}\left[f_{-\omega,k}+(1-e^{iu\omega})n_{\omega}N_{\omega,k} \right],\\
  \Phi_e &= \int\frac{\d\omega \d k}{(2\pi)^2}e^{-\omega \tau + i k \theta}\lambda_{\omega,k}(1-e^{iu\omega})n_{\omega}N_{\omega,k},
\end{align}
\end{widetext}
with $n_\omega := (1-e^{-\beta \omega})^{-1}$.
To check if the junction conditions are satisfied, for example at $t_f=0$ and $\tau=0$, we recognize
\begin{align}
  0 &= \int\frac{\d\omega \d k}{(2\pi)^2}e^{-i\omega t + i k \theta}\lambda_{\omega,k} f_{-\omega,k} \nonumber\\
  &= \int\frac{\d\omega \d k}{(2\pi)^2}e^{-i\omega t + i k \theta}\omega\;\lambda_{\omega,k} f_{-\omega,k}
  \quad
  (t=0).
\end{align}
This follows from the fact that the support of $\lambda$ lies at $t>0$ (see around (4.4.35) of \cite{Skenderis:2008dg}).
This is physically because $f_{-\omega,k}$ propagates the influence of the source $\lambda$ only toward the future since it is the ingoing mode.
A similar relation holds for $f_{\omega,k}$ at $t = u+v$.

Substituting the obtained solution into the action $iS_f - iS_b - S_e$ (with the counterterms in \cite{deHaro:2000vlm}) yields $G(u)$.
Using the equations of motion, only boundary terms survive, and thus we obtain
\begin{widetext}
\begin{align}
  G(u) &= \exp\!\left[-4(\Delta-1)\int\frac{\d\omega \d k}{(2\pi)^2}|\lambda_{\omega,k}|^2O_{\omega,k}\sin\frac{u\omega}{2}\left(\cos\frac{u\omega}{2} + i\sin\frac{u\omega}{2}\coth\frac{\beta\omega}{2} \right) \right],\\
  O_{\omega,k} &:= -r_+^{2\Delta-2}\frac{\Gamma(2-\Delta)\Gamma(\gamma_{\omega,k})\Gamma(\gamma_{\omega,-k})}{\Gamma(\Delta)\Gamma(1-\bar \gamma_{\omega,k})\Gamma(1-\bar \gamma_{\omega,-k})},
\end{align}
\end{widetext}
where $\bar \gamma$ is the complex conjugate of $\gamma$.
Therefore, expanding this to $\mathcal{O}(\lambda^2)$ and Fourier transforming it, we recover \eqref{eq: ex p}.

\bibliography{ref}

\begin{thebibliography}{10}

\bibitem{BardeenCarterHawking}
J.~M. Bardeen, B.~Carter, and S.~W. Hawking.
\newblock The four laws of black hole mechanics.
\newblock {\em Communications in Mathematical Physics}, 31(2):161--170, 1973.

\bibitem{HawkingHartle1st}
S.~W. Hawking and J.~B. Hartle.
\newblock Energy and angular momentum flow into a black hole.
\newblock {\em Communications in Mathematical Physics}, 27(4):283--290, 1972.

\bibitem{Wald:1993nt}
Robert~M. Wald.
\newblock {Black hole entropy is the Noether charge}.
\newblock {\em Phys. Rev. D}, 48(8):R3427--R3431, 1993.
\newblock \href {https://arxiv.org/abs/gr-qc/9307038} {\path{arXiv:gr-qc/9307038}}, \href {https://doi.org/10.1103/PhysRevD.48.R3427} {\path{doi:10.1103/PhysRevD.48.R3427}}.

\bibitem{Bekenstein1}
J.~D. Bekenstein.
\newblock Black holes and the second law.
\newblock {\em Lettere al Nuovo Cimento (1971-1985)}, 4(15):737--740, 1972.

\bibitem{Hawking2nd}
S.~W. Hawking.
\newblock Black holes in general relativity.
\newblock {\em Communications in Mathematical Physics}, 25(2):152--166, 1972.

\bibitem{Bekenstein:1973ur}
Jacob~D. Bekenstein.
\newblock {Black holes and entropy}.
\newblock {\em Phys. Rev. D}, 7:2333--2346, 1973.
\newblock \href {https://doi.org/10.1103/PhysRevD.7.2333} {\path{doi:10.1103/PhysRevD.7.2333}}.

\bibitem{BekensteinGSL}
Jacob~D. Bekenstein.
\newblock Generalized second law of thermodynamics in black-hole physics.
\newblock {\em Phys. Rev. D}, 9:3292--3300, Jun 1974.
\newblock URL: \url{https://link.aps.org/doi/10.1103/PhysRevD.9.3292}, \href {https://doi.org/10.1103/PhysRevD.9.3292} {\path{doi:10.1103/PhysRevD.9.3292}}.

\bibitem{Wall:2009wm}
Aron~C. Wall.
\newblock {Ten Proofs of the Generalized Second Law}.
\newblock {\em JHEP}, 06:021, 2009.
\newblock \href {https://arxiv.org/abs/0901.3865} {\path{arXiv:0901.3865}}, \href {https://doi.org/10.1088/1126-6708/2009/06/021} {\path{doi:10.1088/1126-6708/2009/06/021}}.

\bibitem{Carlip:2014pma}
S.~Carlip.
\newblock {Black Hole Thermodynamics}.
\newblock {\em Int. J. Mod. Phys. D}, 23:1430023, 2014.
\newblock \href {https://arxiv.org/abs/1410.1486} {\path{arXiv:1410.1486}}, \href {https://doi.org/10.1142/S0218271814300237} {\path{doi:10.1142/S0218271814300237}}.

\bibitem{Wall:2018ydq}
Aron~C. Wall.
\newblock {A Survey of Black Hole Thermodynamics}.
\newblock 4 2018.
\newblock \href {https://arxiv.org/abs/1804.10610} {\path{arXiv:1804.10610}}.

\bibitem{Sarkar:2019xfd}
Sudipta Sarkar.
\newblock {Black Hole Thermodynamics: General Relativity and Beyond}.
\newblock {\em Gen. Rel. Grav.}, 51(5):63, 2019.
\newblock \href {https://arxiv.org/abs/1905.04466} {\path{arXiv:1905.04466}}, \href {https://doi.org/10.1007/s10714-019-2545-y} {\path{doi:10.1007/s10714-019-2545-y}}.

\bibitem{Strominger:1996sh}
Andrew Strominger and Cumrun Vafa.
\newblock {Microscopic origin of the Bekenstein-Hawking entropy}.
\newblock {\em Phys. Lett. B}, 379:99--104, 1996.
\newblock \href {https://arxiv.org/abs/hep-th/9601029} {\path{arXiv:hep-th/9601029}}, \href {https://doi.org/10.1016/0370-2693(96)00345-0} {\path{doi:10.1016/0370-2693(96)00345-0}}.

\bibitem{Jacobson:1995ab}
Ted Jacobson.
\newblock {Thermodynamics of space-time: The Einstein equation of state}.
\newblock {\em Phys. Rev. Lett.}, 75:1260--1263, 1995.
\newblock \href {https://arxiv.org/abs/gr-qc/9504004} {\path{arXiv:gr-qc/9504004}}, \href {https://doi.org/10.1103/PhysRevLett.75.1260} {\path{doi:10.1103/PhysRevLett.75.1260}}.

\bibitem{jarzynski2000hamiltonian}
Christopher Jarzynski.
\newblock Hamiltonian derivation of a detailed fluctuation theorem.
\newblock {\em Journal of Statistical Physics}, 98:77--102, 2000.

\bibitem{CrooksFT}
Gavin~E. Crooks.
\newblock Entropy production fluctuation theorem and the nonequilibrium work relation for free energy differences.
\newblock {\em Phys. Rev. E}, 60:2721--2726, Sep 1999.
\newblock URL: \url{https://link.aps.org/doi/10.1103/PhysRevE.60.2721}, \href {https://doi.org/10.1103/PhysRevE.60.2721} {\path{doi:10.1103/PhysRevE.60.2721}}.

\bibitem{Kurchan:2000rzb}
Jorge Kurchan.
\newblock {A Quantum Fluctuation Theorem}.
\newblock 7 2000.
\newblock \href {https://arxiv.org/abs/cond-mat/0007360} {\path{arXiv:cond-mat/0007360}}.

\bibitem{Tasaki2000jarzynski}
Hal Tasaki.
\newblock Jarzynski relations for quantum systems and some applications.
\newblock {\em arXiv preprint cond-mat/0009244}, 2000.

\bibitem{Massar:1999wg}
S.~Massar and R.~Parentani.
\newblock {How the change in horizon area drives black hole evaporation}.
\newblock {\em Nucl. Phys. B}, 575:333--356, 2000.
\newblock \href {https://arxiv.org/abs/gr-qc/9903027} {\path{arXiv:gr-qc/9903027}}, \href {https://doi.org/10.1016/S0550-3213(00)00067-5} {\path{doi:10.1016/S0550-3213(00)00067-5}}.

\bibitem{Iso:2010tz}
Satoshi Iso, Susumu Okazawa, and Sen Zhang.
\newblock {Non-equilibrium fluctuations of black hole horizons and the generalized second law}.
\newblock {\em Phys. Lett. B}, 705:152--156, 2011.
\newblock \href {https://arxiv.org/abs/1008.1184} {\path{arXiv:1008.1184}}, \href {https://doi.org/10.1016/j.physletb.2011.09.114} {\path{doi:10.1016/j.physletb.2011.09.114}}.

\bibitem{Iso:2011gb}
Satoshi Iso and Susumu Okazawa.
\newblock {Stochastic Equations in Black Hole Backgrounds and Non-equilibrium Fluctuation Theorems}.
\newblock {\em Nucl. Phys. B}, 851:380--419, 2011.
\newblock \href {https://arxiv.org/abs/1104.2461} {\path{arXiv:1104.2461}}, \href {https://doi.org/10.1016/j.nuclphysb.2011.05.021} {\path{doi:10.1016/j.nuclphysb.2011.05.021}}.

\bibitem{Liu:2014hna}
Nana Liu, John Goold, Ivette Fuentes, Vlatko Vedral, Kavan Modi, and David~Edward Bruschi.
\newblock {Quantum thermodynamics for a model of an expanding universe}.
\newblock {\em Class. Quant. Grav.}, 33:035003, 2016.
\newblock \href {https://arxiv.org/abs/1409.5283} {\path{arXiv:1409.5283}}, \href {https://doi.org/10.1088/0264-9381/33/3/035003} {\path{doi:10.1088/0264-9381/33/3/035003}}.

\bibitem{Basso:2023fua}
Marcos L.~W. Basso, Jonas Maziero, and Lucas~C. C{\'e}leri.
\newblock {The irreversibility of relativistic time-dilation}.
\newblock {\em Class. Quant. Grav.}, 40(19):195001, 2023.
\newblock \href {https://arxiv.org/abs/2307.12778} {\path{arXiv:2307.12778}}, \href {https://doi.org/10.1088/1361-6382/acf089} {\path{doi:10.1088/1361-6382/acf089}}.

\bibitem{Basso:2024yln}
Marcos L.~W. Basso, Jonas Maziero, and Lucas~Chibebe C{\'e}leri.
\newblock {Quantum Detailed Fluctuation Theorem in Curved Spacetimes: The Observer Dependent Nature of Entropy Production}.
\newblock {\em Phys. Rev. Lett.}, 134(5):050406, 2025.
\newblock \href {https://arxiv.org/abs/2405.03902} {\path{arXiv:2405.03902}}, \href {https://doi.org/10.1103/PhysRevLett.134.050406} {\path{doi:10.1103/PhysRevLett.134.050406}}.

\bibitem{Cai:2024bgh}
Yifan Cai, Tao Wang, and Liu Zhao.
\newblock {General relativistic fluctuation theorems}.
\newblock {\em Phys. Lett. B}, 860:139220, 2025.
\newblock \href {https://arxiv.org/abs/2407.09912} {\path{arXiv:2407.09912}}, \href {https://doi.org/10.1016/j.physletb.2024.139220} {\path{doi:10.1016/j.physletb.2024.139220}}.

\bibitem{Costa:2025vqr}
Rafael L.~S. Costa, Marcos L.~W. Basso, Jonas Maziero, and Lucas~C. C{\'e}leri.
\newblock {Work distribution of quantum fields in static curved spacetimes}.
\newblock 10 2025.
\newblock \href {https://arxiv.org/abs/2510.08265} {\path{arXiv:2510.08265}}.

\bibitem{Cirafici:2024ccs}
Michele Cirafici.
\newblock {Fluctuation theorems, quantum channels and gravitational algebras}.
\newblock {\em JHEP}, 11:089, 2024.
\newblock \href {https://arxiv.org/abs/2408.04219} {\path{arXiv:2408.04219}}, \href {https://doi.org/10.1007/JHEP11(2024)089} {\path{doi:10.1007/JHEP11(2024)089}}.

\bibitem{Cirafici:2024itu}
Michele Cirafici.
\newblock {Gravitational Algebras and Applications to Nonequilibrium Physics}.
\newblock {\em Universe}, 11(1):24, 2025.
\newblock \href {https://arxiv.org/abs/2412.17674} {\path{arXiv:2412.17674}}, \href {https://doi.org/10.3390/universe11010024} {\path{doi:10.3390/universe11010024}}.

\bibitem{Pourhassan:2021mhb}
Behnam Pourhassan, Salman~Sajad Wani, Saheb Soroushfar, and Mir Faizal.
\newblock {Quantum work and information geometry of a quantum Myers-Perry black hole}.
\newblock {\em JHEP}, 10:027, 2021.
\newblock \href {https://arxiv.org/abs/2102.03296} {\path{arXiv:2102.03296}}, \href {https://doi.org/10.1007/JHEP10(2021)027} {\path{doi:10.1007/JHEP10(2021)027}}.

\bibitem{Pourhassan:2022sfk}
Behnam Pourhassan, Houcine Aounallah, Mir Faizal, Sudhaker Upadhyay, Saheb Soroushfar, Yermek~O. Aitenov, and Salman~Sajad Wani.
\newblock {Quantum thermodynamics of an M2-M5 brane system}.
\newblock {\em JHEP}, 05:030, 2022.
\newblock \href {https://arxiv.org/abs/2201.11073} {\path{arXiv:2201.11073}}, \href {https://doi.org/10.1007/JHEP05(2022)030} {\path{doi:10.1007/JHEP05(2022)030}}.

\bibitem{Pourhassan:2022opb}
Behnam Pourhassan, Mahdi Atashi, Houcine Aounallah, Salman~Sajad Wani, Mir Faizal, and Barun Majumder.
\newblock {Quantum thermodynamics of a quantum sized AdS black hole}.
\newblock {\em Nucl. Phys. B}, 980:115842, 2022.
\newblock \href {https://arxiv.org/abs/2205.13584} {\path{arXiv:2205.13584}}, \href {https://doi.org/10.1016/j.nuclphysb.2022.115842} {\path{doi:10.1016/j.nuclphysb.2022.115842}}.

\bibitem{Pourhassan:2022irk}
Behnam Pourhassan, Izzet Sakalli, Xiaoping Shi, Mir Faizal, and Salman~Sajad Wani.
\newblock {Quantum Thermodynamics of an $\alpha^{\prime}$-Corrected Reissner-Nordstr{\"o}m Black Hole}.
\newblock {\em EPL}, 144(2):29001, 2023.
\newblock \href {https://arxiv.org/abs/2301.00687} {\path{arXiv:2301.00687}}, \href {https://doi.org/10.1209/0295-5075/acfff0} {\path{doi:10.1209/0295-5075/acfff0}}.

\bibitem{Masood:2024yio}
Syed Masood.
\newblock {Thermodynamic phase description of charged 4D Gauss-Bonnet black holes in quantum regime}.
\newblock 6 2024.
\newblock \href {https://arxiv.org/abs/2406.05820} {\path{arXiv:2406.05820}}.

\bibitem{Maldacena:1997re}
Juan~Martin Maldacena.
\newblock {The Large N limit of superconformal field theories and supergravity}.
\newblock {\em Adv. Theor. Math. Phys.}, 2:231--252, 1998.
\newblock \href {https://arxiv.org/abs/hep-th/9711200} {\path{arXiv:hep-th/9711200}}, \href {https://doi.org/10.1023/A:1026654312961} {\path{doi:10.1023/A:1026654312961}}.

\bibitem{Gubser:1998bc}
S.~S. Gubser, Igor~R. Klebanov, and Alexander~M. Polyakov.
\newblock {Gauge theory correlators from noncritical string theory}.
\newblock {\em Phys. Lett. B}, 428:105--114, 1998.
\newblock \href {https://arxiv.org/abs/hep-th/9802109} {\path{arXiv:hep-th/9802109}}, \href {https://doi.org/10.1016/S0370-2693(98)00377-3} {\path{doi:10.1016/S0370-2693(98)00377-3}}.

\bibitem{Witten:1998qj}
Edward Witten.
\newblock {Anti-de Sitter space and holography}.
\newblock {\em Adv. Theor. Math. Phys.}, 2:253--291, 1998.
\newblock \href {https://arxiv.org/abs/hep-th/9802150} {\path{arXiv:hep-th/9802150}}, \href {https://doi.org/10.4310/ATMP.1998.v2.n2.a2} {\path{doi:10.4310/ATMP.1998.v2.n2.a2}}.

\bibitem{Engelhardt:2017aux}
Netta Engelhardt and Aron~C. Wall.
\newblock {Decoding the Apparent Horizon: Coarse-Grained Holographic Entropy}.
\newblock {\em Phys. Rev. Lett.}, 121(21):211301, 2018.
\newblock \href {https://arxiv.org/abs/1706.02038} {\path{arXiv:1706.02038}}, \href {https://doi.org/10.1103/PhysRevLett.121.211301} {\path{doi:10.1103/PhysRevLett.121.211301}}.

\bibitem{Engelhardt:2018kcs}
Netta Engelhardt and Aron~C. Wall.
\newblock {Coarse Graining Holographic Black Holes}.
\newblock {\em JHEP}, 05:160, 2019.
\newblock \href {https://arxiv.org/abs/1806.01281} {\path{arXiv:1806.01281}}, \href {https://doi.org/10.1007/JHEP05(2019)160} {\path{doi:10.1007/JHEP05(2019)160}}.

\bibitem{Chandrasekaran:2022eqq}
Venkatesa Chandrasekaran, Geoff Penington, and Edward Witten.
\newblock {Large N algebras and generalized entropy}.
\newblock {\em JHEP}, 04:009, 2023.
\newblock \href {https://arxiv.org/abs/2209.10454} {\path{arXiv:2209.10454}}, \href {https://doi.org/10.1007/JHEP04(2023)009} {\path{doi:10.1007/JHEP04(2023)009}}.

\bibitem{Takeda:2024qbq}
Daichi Takeda.
\newblock {Coarse-graining black holes out of equilibrium with boundary observables on time slice}.
\newblock {\em JHEP}, 05:319, 2024.
\newblock [Erratum: JHEP 10, 154 (2024)].
\newblock \href {https://arxiv.org/abs/2403.07275} {\path{arXiv:2403.07275}}, \href {https://doi.org/10.1007/JHEP05(2024)319} {\path{doi:10.1007/JHEP05(2024)319}}.

\bibitem{Shigemura:2024yeb}
Tomohiro Shigemura, Keito Shimizu, Sotaro Sugishita, Daichi Takeda, and Takuya Yoda.
\newblock {Heat and work in black hole thermodynamics via holography}.
\newblock {\em JHEP}, 05:069, 2025.
\newblock \href {https://arxiv.org/abs/2412.15697} {\path{arXiv:2412.15697}}, \href {https://doi.org/10.1007/JHEP05(2025)069} {\path{doi:10.1007/JHEP05(2025)069}}.

\bibitem{Skenderis:2008dh}
Kostas Skenderis and Balt~C. van Rees.
\newblock {Real-time gauge/gravity duality}.
\newblock {\em Phys. Rev. Lett.}, 101:081601, 2008.
\newblock \href {https://arxiv.org/abs/0805.0150} {\path{arXiv:0805.0150}}, \href {https://doi.org/10.1103/PhysRevLett.101.081601} {\path{doi:10.1103/PhysRevLett.101.081601}}.

\bibitem{Skenderis:2008dg}
Kostas Skenderis and Balt~C. van Rees.
\newblock {Real-time gauge/gravity duality: Prescription, Renormalization and Examples}.
\newblock {\em JHEP}, 05:085, 2009.
\newblock \href {https://arxiv.org/abs/0812.2909} {\path{arXiv:0812.2909}}, \href {https://doi.org/10.1088/1126-6708/2009/05/085} {\path{doi:10.1088/1126-6708/2009/05/085}}.

\bibitem{Glorioso:2018mmw}
Paolo Glorioso, Michael Crossley, and Hong Liu.
\newblock {A prescription for holographic Schwinger-Keldysh contour in non-equilibrium systems}.
\newblock 12 2018.
\newblock \href {https://arxiv.org/abs/1812.08785} {\path{arXiv:1812.08785}}.

\bibitem{Talkner2007fluctuation}
Peter Talkner, Eric Lutz, and Peter H{\"a}nggi.
\newblock Fluctuation theorems: Work is not an observable.
\newblock {\em Physical Review E—Statistical, Nonlinear, and Soft Matter Physics}, 75(5):050102, 2007.

\bibitem{Talkner2007tasaki}
Peter Talkner and Peter H{\"a}nggi.
\newblock The tasaki--crooks quantum fluctuation theorem.
\newblock {\em Journal of Physics A: Mathematical and Theoretical}, 40(26):F569, 2007.

\bibitem{Andrieux2008quantum}
David Andrieux and Pierre Gaspard.
\newblock Quantum work relations and response theory.
\newblock {\em Physical review letters}, 100(23):230404, 2008.

\bibitem{Jarzynski1997nonequilibrium}
Christopher Jarzynski.
\newblock Nonequilibrium equality for free energy differences.
\newblock {\em Physical Review Letters}, 78(14):2690, 1997.

\bibitem{Ortega:2019etm}
Alvaro Ortega, Emma McKay, {\'A}lvaro~M. Alhambra, and Eduardo Mart{\'\i}n-Mart{\'\i}nez.
\newblock {Work distributions on quantum fields}.
\newblock {\em Phys. Rev. Lett.}, 122(24):240604, 2019.
\newblock \href {https://arxiv.org/abs/1902.03258} {\path{arXiv:1902.03258}}, \href {https://doi.org/10.1103/PhysRevLett.122.240604} {\path{doi:10.1103/PhysRevLett.122.240604}}.

\bibitem{DornerQuantumWork}
R.~Dorner, S.~R. Clark, L.~Heaney, R.~Fazio, J.~Goold, and V.~Vedral.
\newblock Extracting quantum work statistics and fluctuation theorems by single-qubit interferometry.
\newblock {\em Phys. Rev. Lett.}, 110:230601, Jun 2013.
\newblock URL: \url{https://link.aps.org/doi/10.1103/PhysRevLett.110.230601}, \href {https://doi.org/10.1103/PhysRevLett.110.230601} {\path{doi:10.1103/PhysRevLett.110.230601}}.

\bibitem{Mazzola:2014dvw}
Laura Mazzola, Gabriele De~Chiara, and Mauro Paternostro.
\newblock {Detecting the work statistics through Ramsey-like interferometry}.
\newblock {\em Int. J. Quant. Inf.}, 12(02):1461007, 2014.
\newblock \href {https://arxiv.org/abs/1401.0566} {\path{arXiv:1401.0566}}, \href {https://doi.org/10.1142/s0219749914610073} {\path{doi:10.1142/s0219749914610073}}.

\bibitem{vanRees:2009rw}
Balt~C. van Rees.
\newblock {Real-time gauge/gravity duality and ingoing boundary conditions}.
\newblock {\em Nucl. Phys. B Proc. Suppl.}, 192-193:193--196, 2009.
\newblock \href {https://arxiv.org/abs/0902.4010} {\path{arXiv:0902.4010}}, \href {https://doi.org/10.1016/j.nuclphysbps.2009.07.078} {\path{doi:10.1016/j.nuclphysbps.2009.07.078}}.

\bibitem{Brown:1992br}
J.~David Brown and James~W. York, Jr.
\newblock {Quasilocal energy and conserved charges derived from the gravitational action}.
\newblock {\em Phys. Rev. D}, 47:1407--1419, 1993.
\newblock \href {https://arxiv.org/abs/gr-qc/9209012} {\path{arXiv:gr-qc/9209012}}, \href {https://doi.org/10.1103/PhysRevD.47.1407} {\path{doi:10.1103/PhysRevD.47.1407}}.

\bibitem{Balasubramanian:1999re}
Vijay Balasubramanian and Per Kraus.
\newblock {A Stress tensor for Anti-de Sitter gravity}.
\newblock {\em Commun. Math. Phys.}, 208:413--428, 1999.
\newblock \href {https://arxiv.org/abs/hep-th/9902121} {\path{arXiv:hep-th/9902121}}, \href {https://doi.org/10.1007/s002200050764} {\path{doi:10.1007/s002200050764}}.

\bibitem{Hartle:1976tp}
J.~B. Hartle and S.~W. Hawking.
\newblock {Path Integral Derivation of Black Hole Radiance}.
\newblock {\em Phys. Rev. D}, 13:2188--2203, 1976.
\newblock \href {https://doi.org/10.1103/PhysRevD.13.2188} {\path{doi:10.1103/PhysRevD.13.2188}}.

\bibitem{deHaro:2000vlm}
Sebastian de~Haro, Sergey~N. Solodukhin, and Kostas Skenderis.
\newblock {Holographic reconstruction of space-time and renormalization in the AdS / CFT correspondence}.
\newblock {\em Commun. Math. Phys.}, 217:595--622, 2001.
\newblock \href {https://arxiv.org/abs/hep-th/0002230} {\path{arXiv:hep-th/0002230}}, \href {https://doi.org/10.1007/s002200100381} {\path{doi:10.1007/s002200100381}}.

\bibitem{Aharony:2001pa}
Ofer Aharony, Micha Berkooz, and Eva Silverstein.
\newblock {Multiple trace operators and nonlocal string theories}.
\newblock {\em JHEP}, 08:006, 2001.
\newblock \href {https://arxiv.org/abs/hep-th/0105309} {\path{arXiv:hep-th/0105309}}, \href {https://doi.org/10.1088/1126-6708/2001/08/006} {\path{doi:10.1088/1126-6708/2001/08/006}}.

\bibitem{Witten:2001ua}
Edward Witten.
\newblock {Multitrace operators, boundary conditions, and AdS / CFT correspondence}.
\newblock 12 2001.
\newblock \href {https://arxiv.org/abs/hep-th/0112258} {\path{arXiv:hep-th/0112258}}.

\bibitem{Berkooz:2002ug}
Micha Berkooz, Amit Sever, and Assaf Shomer.
\newblock {'Double trace' deformations, boundary conditions and space-time singularities}.
\newblock {\em JHEP}, 05:034, 2002.
\newblock \href {https://arxiv.org/abs/hep-th/0112264} {\path{arXiv:hep-th/0112264}}, \href {https://doi.org/10.1088/1126-6708/2002/05/034} {\path{doi:10.1088/1126-6708/2002/05/034}}.

\bibitem{Sever:2002fk}
Amit Sever and Assaf Shomer.
\newblock {A Note on multitrace deformations and AdS/CFT}.
\newblock {\em JHEP}, 07:027, 2002.
\newblock \href {https://arxiv.org/abs/hep-th/0203168} {\path{arXiv:hep-th/0203168}}, \href {https://doi.org/10.1088/1126-6708/2002/07/027} {\path{doi:10.1088/1126-6708/2002/07/027}}.

\bibitem{Aharony:2005sh}
Ofer Aharony, Micha Berkooz, and Boaz Katz.
\newblock {Non-local effects of multi-trace deformations in the AdS/CFT correspondence}.
\newblock {\em JHEP}, 10:097, 2005.
\newblock \href {https://arxiv.org/abs/hep-th/0504177} {\path{arXiv:hep-th/0504177}}, \href {https://doi.org/10.1088/1126-6708/2005/10/097} {\path{doi:10.1088/1126-6708/2005/10/097}}.

\bibitem{Aharony:2006hz}
Ofer Aharony, Adam~B. Clark, and Andreas Karch.
\newblock {The CFT/AdS correspondence, massive gravitons and a connectivity index conjecture}.
\newblock {\em Phys. Rev. D}, 74:086006, 2006.
\newblock \href {https://arxiv.org/abs/hep-th/0608089} {\path{arXiv:hep-th/0608089}}, \href {https://doi.org/10.1103/PhysRevD.74.086006} {\path{doi:10.1103/PhysRevD.74.086006}}.

\bibitem{Shiraishi2023introduction}
Naoto Shiraishi.
\newblock An introduction to stochastic thermodynamics.
\newblock {\em Fundamental Theories of Physics. Springer, Singapore}, 2023.

\bibitem{Esposito2009nonequilibrium}
Massimiliano Esposito, Upendra Harbola, and Shaul Mukamel.
\newblock Nonequilibrium fluctuations, fluctuation theorems, and counting statistics in quantum systems.
\newblock {\em Reviews of modern physics}, 81(4):1665--1702, 2009.

\bibitem{Karch:2023wui}
Andreas Karch, Mianqi Wang, and Merna Youssef.
\newblock {AdS Higgs mechanism from double trace deformed CFT}.
\newblock {\em JHEP}, 02:044, 2024.
\newblock \href {https://arxiv.org/abs/2311.10135} {\path{arXiv:2311.10135}}, \href {https://doi.org/10.1007/JHEP02(2024)044} {\path{doi:10.1007/JHEP02(2024)044}}.

\bibitem{Geng:2023ynk}
Hao Geng.
\newblock {Open AdS/CFT via a double-trace deformation}.
\newblock {\em JHEP}, 09:012, 2024.
\newblock \href {https://arxiv.org/abs/2311.13633} {\path{arXiv:2311.13633}}, \href {https://doi.org/10.1007/JHEP09(2024)012} {\path{doi:10.1007/JHEP09(2024)012}}.

\bibitem{Karch:2025hof}
Andreas Karch and Merna Youssef.
\newblock {Dissipation in Open Holography}.
\newblock 9 2025.
\newblock \href {https://arxiv.org/abs/2509.14312} {\path{arXiv:2509.14312}}.

\bibitem{Ishii:2025qpy}
Takanori Ishii and Daichi Takeda.
\newblock {Lindblad dynamics in holography}.
\newblock {\em Phys. Rev. D}, 112(4):046020, 2025.
\newblock \href {https://arxiv.org/abs/2504.17320} {\path{arXiv:2504.17320}}, \href {https://doi.org/10.1103/l2rx-t9dd} {\path{doi:10.1103/l2rx-t9dd}}.

\end{thebibliography}
\end{document}